\def\hcorrection#1{\advance\hoffset by #1 }
\def\vcorrection#1{\advance\voffset by #1 }

\documentstyle[11pt]{article}

\vcorrection{-0.70in}
\hcorrection{-0.60in}
\begin{document}

\title{On pressure and velocity flow boundary  conditions
for the lattice Boltzmann BGK model}
\author{
Qisu Zou \thanks{%
 Theoretical Division, Los Alamos National Lab, Los Alamos, NM 87545}
\thanks{%
 Dept. of Math., Kansas State University, Manhattan, KS 66506}
and Xiaoyi He \thanks{%
Center for Nonlinear Studies, Los Alamos National Lab } \thanks{%
Theoretical Biology and Biophysics Group, Los Alamos National Lab}
}
\date{}
\maketitle

\vspace{7mm}



\vspace{7mm}

\begin{abstract}
Pressure (density) and
velocity boundary conditions inside a flow domain are studied
for 2-D and 3-D lattice Boltzmann BGK models (LBGK)
and new method to specify these
conditions are proposed. These conditions are consistent with
the boundary condition we proposed in \cite{he2} using an idea of
bounce-back of non-equilibrium
distribution.
These  conditions give excellent results
for the regular LBGK
models, and were shown to be second-order accurate by numerical examples.
When they are used together with the improved incompressible LBGK model
in \cite{zoui},
the  simulation results recover
the analytical solution of the plane Poiseuille flow
driven by pressure (density) difference
with machine accuracy.

\end{abstract}


\section{Introduction}

The lattice Boltzmann equation (LBE) method has achieved great success
for simulation of transport phenomena in recent years.
Among different LBE methods, the lattice Boltzmann BGK model is considered
more robust \cite{sdqo}.
Besides,  theoretical discussion is easier for the  LBGK due to its simple
form.
Some recent theoretical discussions on LBGK \cite{he2,zou,he1}
have enhanced our understanding of the method  and the effect of boundary
conditions.
In \cite{zou}, analytical solutions of distribution functions
for plane Poiseuille flow with forcing
 and plane Couette flow
have been obtained for the 2-D triangular and square
lattice Boltzmann BGK models. It is found that the bounce-back boundary
condition produces
distribution functions with a first-order error compared with the
analytical distribution functions.
In \cite{he1},
a new technique was developed to seek the analytic solution of
LBGK model for some simple flow.
For example, the velocity profile from the
2-D square and triangular LBGK models are shown to satisfy a second-order
difference equation of the Navier-Stokes equation in the case of plane
Poiseuille flow with forcing and  Couette flow.
The technique is generalized in \cite{he2}
to include steady-state flows with both $x$ and $y$ velocities,
which are  assumed to be independent of $x$.
The analysis provides a framework to analyze any velocity
boundary condition. For example, the analysis explains why  the
velocity boundary condition for the 2-D triangular LBGK
model proposed in \cite{noble}
generates results of machine accuracy
for plane Poiseuille flow with forcing.

In practice, however, a flow is often  driven  by pressure difference.
In general, the pressure gradient through the
flow field is not a constant and the local pressure gradient is unknown
before solving the flow. Hence the pressure gradient in many cases cannot
be replaced by an external force in LBGK computations.
In this situation, boundary conditions usually need be implemented by
giving prescribed pressure or velocity on some ``flow boundaries'',
which are not solid walls  or interfaces of two distinct fluids.
Instead, they are imaginary boundaries inside a flow domain
(e.g. inlet and outlet in a pipe flow). Their existence is purely for the
convenience of study.
The implementation of these  boundary conditions in LBGK
is very important but it has not yet been well studied.

Since in lattice Boltzmann method, the pressure is related
to the density  by the isothermal
equation of state as   $p = c_s^2 \rho$ ($c_s$ is the sound speed of the
model), a specification of pressure difference amounts to  a specification
of density difference.
 Early works (see, for example, \cite{daryl})
to implement pressure (density) flow boundary  condition is simply
to assign the equilibrium distribution computed with the specified
density and some velocity (maybe zero)
to the distribution function.
This method introduces significant errors: the real pressure gradient obtained
in
the simulation for the Poiseuille flow is not a constant. It is approximately
a constant only some distance away from  the inlet and outlet of the channel.
Besides, even away from the inlet and outlet region,
the pressure gradient is different from the intended value.
Maier {\it et al.} \cite{bob} proposed  an alternative
pressure or velocity flow boundary condition
for the  3-D 15-velocity direction LBGK model,
and their results are greatly improved over the equilibrium distribution
approach.
The pressure or velocity flow boundary condition in \cite{bob}
is obtained through a post-streaming rule to the distribution functions
based on an extrapolation.
However,
this pressure or velocity boundary condition is still to be improved
due to some
inconsistency (see discussion in Section 2).
Its inaccuracy is most noticeable in the following case:
when this pressure or velocity boundary condition is applied to
the modified LBGK \cite{zoui},
which corresponds to a macroscopic momentum equation
having the analytical solution of Poiseuille flow with pressure (density)
gradient, the simulation results are far from the analytical results.

In this paper, we propose a general way to specify
pressure or velocity on flow boundaries.
The implementation is a natural extension of the wall boundary condition
described in our previous paper \cite{he2}.
The result shows a clear improvement to the flow boundary
conditions in \cite{bob} for ordinary LBGK models.
Besides, for the  modified LBGK model,
These flow boundary conditions produce results of machine accuracy for
Poiseuille flow.

\section{Pressure or Velocity Flow Boundary Condition of the 2-D Square
Lattice LBGK Model}

\subsection{Governing Equation}

 The square lattice LBGK model (d2q9) is expressed as
(\cite{ccmm},\cite{qian2},\cite{hchen}):
\begin{equation}
f_i({\bf x}+\delta {\bf e}_i, t+\delta)-f_i({\bf x},t)=
 -\frac 1{\tau}[f_i({\bf x},t)-f_i^{(eq)}({\bf x},t)], \ \ i=0,1,...,8,
\label{eq:lbgk}
\end{equation}
where  the equation is written in physical units.
Both the time step and the lattice spacing have the value of $\delta$
in physical units.
$f_{i}({\bf x},t) $ is the density distribution
function along the direction ${\bf e}_i$ at (${\bf x}, t$).  The
particle speed ${\bf e}_i$'s are given by
 ${\bf e}_{i} = (\cos(\pi(i-1)/2), \sin(\pi(i-1)/2), i = 1,2,3,4$, and
${\bf e}_{i} = \sqrt{2} (\cos(\pi(i-4-\frac{1}{2})/2),
\sin(\pi(i-4-\frac{1}{2})/2), i = 5,6,7,8$.
Rest particles of type 0 with ${\bf e}_{0} = 0$ is also allowed
(see Fig.~1).
The  right hand side represents the collision term
 and $\tau $ is the single relaxation time which controls the rate
of approach to equilibrium.
The density per node, $ \rho $, and the macroscopic flow velocity,
 ${\bf u} = (u_x, u_y)$,
are defined in terms of the particle distribution function by
\begin{equation}
 \sum_{i=0}^8 f_i=\rho,  \ \ \ \
 \sum_{i=1}^8 f_i {\bf e}_i =\rho {\bf u}.
\label{eq:dens}
\end{equation}
The equilibrium distribution functions
$f_i^{(eq)}({\bf x},t) $
depend only on local density and
velocity and they
can be chosen in the following form (the model d2q9 \cite{qian2}):
\begin{eqnarray}
f_{0}^{(eq)}&=&\frac{4}{9}\rho[1-\frac{3}{2}{\bf u}\cdot{\bf u}],
\mbox{\hspace{1.65 in}}
\nonumber\\
f_{i}^{(eq)}&=&\frac{1}{9}\rho[1+3({\bf e}_{i}\cdot
{\bf u})+\frac{9}{2}({\bf e}_{i}\cdot {\bf u})^2-\frac{3}{2}{\bf u}\cdot
{\bf u}],\;\; i =  1,2,3,4 \\
f_{i}^{(eq)}&=&\frac{1}{36}\rho[1+3({\bf e}_{i}\cdot
{\bf u})+\frac{9}{2}({\bf e}_{i}\cdot {\bf u})^2-\frac{3}{2}{\bf u}\cdot
{\bf u}], \;\; i =  5,6,7,8 . \nonumber
\label{eq:equil}
\end{eqnarray}

A Chapman-Enskog procedure can be applied to Eq.~(\ref{eq:lbgk}) to derive
the macroscopic equations  of the model.  They are given by:
the continuity equation (with an error term $O(\delta^2)$ being omitted):
\begin{equation}
\frac{\partial \rho}{\partial t} + \nabla \cdot (\rho {\bf u})=0,
\label{eq:cont}
\end{equation}
and the momentum equation (with terms of  $O(\delta^2)$ and $O(\delta u^3)$
 being omitted):
\begin{equation}
\partial_{t}(\rho u_{\alpha})+\partial_{\beta}(\rho u_{\alpha} u_{\beta})
=
-\partial_{\alpha}(c_s^2 \rho)+\partial_{\beta}(2 \nu \rho S_{\alpha \beta}),
\label{eq:momen}
\end{equation}
where the Einstein  summation convention is used.
$S_{\alpha \beta}=\frac{1}{2}(\partial_{\alpha}u_{\beta}+\partial_{\beta}
u_{\alpha})$ is the strain-rate tensor. The pressure is given by
$p = c_s^2 \rho$, where $c_s$ is the speed of sound with
${\displaystyle c_s^2= \frac{1}{3}, }$
and
${\displaystyle \nu=\frac{2 \tau -1}{6} \delta}$,
with $\nu $ being the  the kinematic viscosity.
The form of the error terms and the derivation of these equations can be found
in \cite{qian3,hou1}.

In this paper, we will take the Poiseuille flow
as an example to study the pressure (density) or velocity inlet/outlet
condition.
The analytical solution of Poiseuille flow
in a channel with width $2 L $ for the Navier-Stokes equation is given by:
\begin{equation}
 u_x = u_0 (1 - \frac{y^2}{L^2}) ,
\;\;\;\; u_y = 0, \;\;\;\;
 \frac{\partial p}{\partial x} = - G, \;\;\;\;
 \frac{\partial p}{\partial y} = 0, \;\;\;\;
\label{eq:pois}
\end{equation}
where the pressure gradient
$G$ is a constant related to the centerline velocity $u_0$ by
\begin{equation}
 G = 2 \rho \nu u_0/L^2 ,
\label{eq:g}
\end{equation}
  and the flow density $\rho$ is a constant. The Reynolds number is defined
as Re = $u_0 (2L)/\nu$.

The Poiseuille flow is an exact solution of the steady-state
incompressible Navier-Stokes equations:
\begin{equation}
 \nabla \cdot  {\bf u}=0.
\label{eq:a64}
\end{equation}
\begin{equation}
\partial_{\beta}( u_{\alpha} u_{\beta})
=- \partial_{\alpha} (\frac{p}{\rho_0}) + \nu \partial_{\beta \beta}
u_{\alpha},
\label{eq:a65}
\end{equation}
On the other hand, the steady-state  macroscopic equations of LBGK model,
Eq.~(\ref{eq:lbgk}), is given by:
\begin{equation}
 \nabla \cdot (\rho {\bf u})=0 ,
\label{eq:conts}
\end{equation}
\begin{equation}
\partial_{\beta}(\rho u_{\alpha} u_{\beta})
=- \partial_{\alpha}(c_s^2 \rho)+\frac{2\tau-1}{6} \delta
\partial_{\beta} \partial_{\beta} (\rho u_{\alpha}) .
  \ \
\label{eq:momens}
\end{equation}
These equations are different from  the incompressible Navier-Stokes equations
Eqs.~(\ref{eq:a64},\ref{eq:a65})
by terms containing  the spatial derivative of $\rho$. These discrepancies
are called  compressibility error in LBE model.
Thus, the Poiseuille flow given by
Eq.~(\ref{eq:pois}) is not the exact solution of Eqs.~(\ref{eq:conts},
\ref{eq:momens}) when pressure (density) gradient drives the flow.
That is, due to change of pressure (density) in the $x-$direction,
$u_x$ is not constant in the $x-$direction, and
the velocity profile of the solution of
 Eqs.~(\ref{eq:conts}, \ref{eq:momens})
is no loger a parabolic profile. For a fixed Mach number
($u_0$ fixed),
  as $\delta \rightarrow 0$, the velocity
of the LBGK simulation  will not converge to the velocity in
Eq.~(\ref{eq:pois}) because the compressibility error becomes
dominating. This phenomenon is seen in the result in \cite{bob},
where the error of velocity increases as the number of lattice grid
increases.
Besides, from $\partial_x (\rho u_x) = 0$ (suppose
$u_y =0$ in the simulation), one can see that
$u_x$ should be increasing linearly along the flow direction since
$\rho$ is decreasing linearly.
This makes the comparison of $u_x$ with the analytical velocity of
Poiseuille flow somehow ambiguous.

To make a more sensible study for Poiseuille flow with pressure (density)
or velocity flow boundary condition, it is better to use the improved
incompressible LBGK model proposed in \cite{zoui}.
The model (called d2q9i)
is given by Eq.~(\ref{eq:lbgk}) with the same ${\bf e}_i$ and
the following equilibrium distributions:
\begin{eqnarray}
 f_{0}^{(eq)}&=&\frac{4}{9} [ \rho -  \frac{3}{2} {\bf v}\cdot {\bf v} ] ,
  \nonumber \\
 f_{i}^{(eq)}&=&\frac{1}{9}  [ \rho + 3 {\bf e}_{i} \cdot {\bf v} +
  \frac{9}{2}({\bf e}_{i} \cdot {\bf v})^2 - \frac{3}{2} {\bf v}\cdot{\bf v} ],
\;
  i=1,2,3,4 , \nonumber \\
 f_{i}^{(eq)}&=&\frac{1}{36} [ \rho + 3 {\bf e}_{i} \cdot {\bf v} +
  \frac{9}{2}({\bf e}_{i} \cdot {\bf v})^2 - \frac{3}{2}{\bf v}\cdot {\bf v} ]
,\; i = 5,6,7,8 ,
\label{eq:eqi}
\end{eqnarray}
and
\begin{equation}
 \sum_{i=0}^8 f_{i} = \sum_{i=0}^8 f_{i}^{(eq)} =\rho,  \ \ \ \
 \sum_{i=1}^8 f_i {\bf e}_{i} =
 \sum_{i=1}^8 f_i^{(eq)} {\bf e}_{i} ={\bf v},
\label{eq:densi}
\end{equation}
where ${\bf v}=(v_x, v_y)$ (like the momentum in the ordinary LBGK model)
is used to represent the flow velocity.
 The macroscopic equations of d2q9i in the steady-state case (apart from error
terms of $O(\delta^2)$:
\begin{equation}
 \nabla \cdot  {\bf v}=0 ,
\label{eq:conti}
\end{equation}
\begin{equation}
\partial_{\beta}( v_{\alpha} v_{\beta})
=- \partial_{\alpha}(c_s^2 \rho)+ \nu
\partial_{\beta \beta} v_{\alpha} ,
\label{eq:momeni}
\end{equation}
are exactly the steady-state incompressible
Navier-Stokes equation. In the model,
pressure is related to density by
$c_s^2 \rho = p/\rho_0$ ($c_s^2 = 1/3$) for a flow with constant density like
Poiseuille flow, and $\nu = \frac{2\tau-1}{6} \delta$.

If the flow is steady, d2q9i is superior to d2q9 to simulating incompressible
flows.  If the flow is unsteady,
one may consider the continuity equation derived from d2q9 given by
$\partial_t \rho + (\nabla \cdot \rho) {\bf u} +
\rho \nabla \cdot {\bf u} = 0$.  In a situation where the first two terms
likely
cancel  each other, d2q9 may be of advantage (to approximate the
continuity equation $\nabla \cdot {\bf u}$). If the first two terms have
the same sign, then d2q9i is better.

\subsection{Review of The Velocity Wall Boundary Condition}

It is proved in \cite{he2,he1} that if the flow is steady and independent of
$x$, then the solution  $f_i$  of
Eq.~(\ref{eq:lbgk})  produces a velocity
profile that satisfies a difference equation which is a second-order
approximation of the Navier-Stokes equation.  If the boundary condition
is chosen correctly,
then the difference equation near the boundary is
consistent with the difference equation inside.

A velocity wall boundary condition is proposed in \cite{he2}
as follows:
take the case of a bottom node in Fig.~1,  the boundary is aligned with
$x-$direction with
$f_4, f_7, f_8$ pointing into the wall.
After streaming, $f_0, f_1, f_3, f_4, f_7, f_8$ are known.
Suppose that $u_x, u_y$ are specified on the wall, we need to
determine $f_2, f_5, f_6$ and $\rho$ from Eqs.~(\ref{eq:dens}),
which can be put into  the form:
\begin{equation}
 f_2+f_5+f_6 =
 \rho - (f_0+f_1+f_3+f_4+f_7+f_8) ,
\label{eq:rho}
\end{equation}
\begin{equation}
 f_5-f_6 = \rho u_x - (f_1-f_3-f_7+f_8) ,  \mbox{\hspace{.9in}}
\label{eq:u0}
\end{equation}
\begin{equation}
 f_2+f_5+f_6 = \rho u_y + (f_4+f_7+f_8) .\mbox{\hspace{.9in}}
\label{eq:v0}
\end{equation}
Consistency of Eqs.~(\ref{eq:rho},\ref{eq:v0}) gives
\begin{equation}
 \rho =  \frac{1}{1-u_y} [f_0+f_1+f_3+2(f_4+f_7+f_8)].
\label{eq:rhod}
\end{equation}
\par
We assume the bounce-back rule is still correct for the
non-equilibrium part of the particle distribution normal to
the boundary (in this case, $f_2 - f_2^{(eq)}=f_4 - f_4^{(eq)}$).
With $f_2$ known, $f_5, f_6$ can be found as
\begin{eqnarray}
f_2&=&f_4+\frac 23 \rho u_y  , \nonumber \\
f_5&=&f_7-\frac 12 (f_1-f_3)+\frac 12  \rho u_x +\frac 16  \rho u_y , \nonumber
 \\
f_6&=&f_8+\frac 12 (f_1-f_3)-\frac 12\rho u_x +\frac 16 \rho u_y.
\label{eq:f256}
\end{eqnarray}
The collision step is applied to the boundary nodes also.
For non-slip boundaries,
this boundary condition is reduced to that in \cite{bob}.

\subsection{Specification of Pressure on a  Flow Boundary }

Now let us turn to pressure (density) flow boundary condition.
Its derivation is based on Eq.~(\ref{eq:dens}) as for velocity wall
boundary condition.
Suppose a flow boundary (take the inlet in Fig.~1 as example)
is along the $y-$direction,
and the pressure is to be specified on it. Suppose that $u_y$ is also
specified (e.g. $u_y=0$ at the inlet in a channel
flow).
After streaming, $f_2,f_3,f_4,f_6,f_7$  are known, $\rho
= \rho_{in}, u_y=0$ are specified
at inlet. We need to  determine
$u_x$ and $ f_1, f_5, f_8$ from Eq.~(\ref{eq:dens}) as following:
\begin{equation}
 f_1+f_5+f_8 =
 \rho_{in} - (f_0+f_2+f_3+f_4+f_6+f_7) ,
\label{eq:rhoi}
\end{equation}
\begin{equation}
 f_1+f_5+f_8 =
  \rho_{in} u_x + (f_3+f_6+f_7) ,  \mbox{\hspace{.9in}}
\label{eq:u0i}
\end{equation}
\begin{equation}
 f_5-f_8 = f_2-f_4+f_6-f_7  .\mbox{\hspace{.9in}}
\label{eq:v0i}
\end{equation}
Consistency of Eqs.~(\ref{eq:rhoi},\ref{eq:u0i}) gives
\begin{equation}
 u_x =  1 - \frac{ [f_0+f_2+f_4+2(f_3+f_6+f_7)]}{\rho_{in}} .
\label{eq:u0d}
\end{equation}
\par
We use  bounce-back rule for the
non-equilibrium part of the particle distribution normal to
the inlet to find $f_1 - f_1^{(eq)}=f_3 - f_3^{(eq)}$.
With $f_1$ known,  $f_5, f_8$ are obtained by the remaining two
equations:
\begin{eqnarray}
f_1&=&f_3+\frac 23 \rho_{in} u_x , \nonumber \\
f_5&=&f_7-\frac 12 (f_2-f_4)+\frac 16  \rho_{in} u_x  ,  \nonumber \\
f_8&=&f_6+\frac 12 (f_2-f_4)+\frac 16 \rho_{in} u_x .
\label{eq:f158}
\end{eqnarray}

The corner node at inlet needs some special treatment. Take the
bottom node at inlet as an example,
after streaming, $ f_3,f_4,f_7$  are known; $\rho$ is specified,
and $u_x=u_y = 0$. We need to determine
$ f_1,f_2,f_5,f_6,f_8$. We use  bounce-back rule for the
non-equilibrium part of the particle distribution normal to
the inlet and the boundary to find:
\begin{equation}
 f_1 = f_3 + (f_1^{(eq)}- f_3^{(eq)}) = f_3 , \;\;
 f_2 = f_4 + (f_1^{(eq)}- f_3^{(eq)})  = f_4,
\label{eq:f13}
\end{equation}
Using these $f_1, f_2$ in Eqs.~(\ref{eq:dens}),
we find:
\begin{equation}
 f_5 = f_7, \;\; f_6=f_8=\frac 12 [\rho_{in}-(f_0+f_1+f_2+f_3+f_4+f_5+f_7)] .
\label{eq:f5678}
\end{equation}
Similar procedure can be applied to top inlet node and outlet nodes
including outlet corner nodes.
The case that $\rho$ and
non-zero $u_y$ is specified at a flow boundary along  $y-$direction
can be handled in the same way.

Here, it is useful to compare our pressure boundary condition
with that proposed in \cite{bob}, which is given by the following
post-streaming rule (an extrapolation) at inlet:
after streaming, $f_1, f_5, f_8$ are calculated as
\begin{equation}
 f_i({\bf x}, t)= f_i^+({\bf x}, t-\delta) -
 (f_j({\bf x}, t)-f_j^+({\bf x}, t-\delta) ) , \;\; i = 1, 5,8
\label{eq:posts}
\end{equation}
where $f_i^+({\bf x}, t-\delta)
 \equiv  f_i({\bf x}, t-\delta) - \frac{1}{\tau}(f_i({\bf x}, t-\delta)
-f_i^{(eq)}({\bf x}, t-\delta))$
is the distribution functions at previous
time step after collision and before streaming, $f_j$ is along ${\bf e}_j$
with ${\bf e}_j = {\bf e}_i - 2 {\bf e}_n$ (the inner normal
${\bf e}_n = {\bf e}_1$ in the case).
Thus, for $i=1,5,8$, $j=3,6,7$ respectively.
The density $\rho $ is set to the specified inlet value and $u_y$ is set to
zero
to compute  $f_i^{(eq)}({\bf x}, t)$.
At the bottom, $f_1, f_8$ are computed using
Eq.~(\ref{eq:posts}) and then $f_2, f_5, f_6$ are obtained  in the
treatment of wall boundary condition.
Notice, however,
that  at the inlet,
$ \sum_{i = 0}^{8} f_i$ may not be  equal to the specified density and
$ \sum_{i = 1}^{8} e_{iy} f_i$ may not be equal to zero with this
post-streaming
operation. This inconsistency
causes some inaccuracy in simulations and leaves room
for improvement.

\subsection{Specification of Velocity on a Flow Boundary }

In some calculations, velocities $u_x, u_y$ are specified at a flow boundary
(take the inlet in Fig.~1 as example).
In the case of channel flow, after streaming, $f_2,f_4,f_3,f_6,f_7$  are known
at inlet.
$u_x, u_y $ are specified at inlet
 (for the special case of Poiseuille flow, $u_y=0$),
we need to determine
$\rho$ and $ f_1, f_5, f_8$.
This is actually equivalent to a velocity wall boundary
condition. Using our velocity wall boundary condition in \cite{he2}
previously described, we find:
\begin{eqnarray}
\rho&=& \frac{1}{1-u_{x}} [f_0+f_2+f_4+2(f_3+f_6+f_7)], \nonumber \\
f_1&=&f_3+\frac 23 \rho u_{x} , \nonumber \\
f_5&=&f_7-\frac 12 (f_2-f_4)+\frac 12 \rho u_y +
 \frac 16  \rho u_{x}  , \nonumber  \\
f_8&=&f_6+\frac 12 (f_2-f_4)-\frac 12 \rho u_y +
 \frac 16 \rho u_{x} .
\label{eq:bc}
\end{eqnarray}
The effect of specifying  velocity at inlet is similar to specifying
pressure (density) at inlet. Density difference in the flow
can be  generated by the velocity inlet condition.

At the inlet bottom (non-slip boundary), special treatment is needed.
After streaming,
$f_1, f_2, f_5, f_6, f_8$ need to be determined. Using bounce-back
on normal distributions gives:
\[ f_1 = f_3, \;\; f_2 = f_4 ,   \]
expressions of $x, y$ momenta give:
\begin{eqnarray}
f_5-f_6+f_8 = -(f_1-f_3-f_7)= f_7 , \nonumber \\
f_5+f_6-f_8 = -(f_2-f_4-f_7)= f_7 ,
\label{eq:veli}
\end{eqnarray}
which gives
\begin{eqnarray}
f_5&=&f_7 ,  \nonumber \\
f_6&=&f_8 = \frac 12 [\rho - (f_0+f_1+f_2+f_3+f_4+f_5+f_7) ] ,
\label{eq:veli1}
\end{eqnarray}
but there is no more equation available to determine $\rho$. The situation
is similar to a corner wall node (the intersection of two perpendicular
walls).
In the situation, $\rho$ at the inlet bottom node can be taken as the
$\rho$ of its neighboring  flow node, thus the velocity inlet condition
is specified.

{}From the discussion given above, we can unify boundary conditions (on a wall
boundary or in a flow boundary) in 2-D
simulation on a straight boundary as:
\begin{itemize}
\item  Given $u_x, u_y$, find $\rho$ and unknown $f_i$'s.
\item  Given $\rho$ and the velocity along the boundary, find the velocity
normal to the boundary and unknown $f_i$'s.
\end{itemize}
Eq.~(\ref{eq:dens}) is used to determine $\rho$ or the normal velocity and
the unknown $f_i$'s. The formula are given in sections 2.2, 2.3, 2.4.

Again, it is useful to compare our
velocity flow boundary condition with that  proposed in \cite{bob},
which is given by the following
post-streaming rule (a
zeroth-order extrapolation) at inlet:
after streaming, $f_1, f_5, f_8$ are calculated  using
\begin{equation}
 f_i({\bf x}, t)= f_i^+({\bf x}, t-\delta) , \;\; i = 1,5,8
\label{eq:postsv}
\end{equation}
where $f_i^+({\bf x}, t-\delta)$ is the distribution function at previous
time step after collision.
After $\rho $ is computed using $f_i$'s,
$f_i^{(eq)}({\bf x}, t)$ can be computed using this $\rho$ and the specified
velocities $u_x, u_y$.
In this approach, the determination of unknown $f_i$'s  does not use the
information of known $f_i$'s at present time. This is inconsistent with
the present distribution in the flow.
Suppose that initially, $f_i^{(eq)},\ i = 0, \cdots, 8$  are computed by using
some density $\rho_0$ and the specified  $u_x, u_y$, and one assigns
$f_i = f_i^{(eq)}, \ i = 0, \cdots, 8$, then collision does not change
$f_i$, and the post-streaming rule Eq.~(\ref{eq:postsv}) does not
change  $f_i$ and $\rho$. Hence  $f_i=f_i^{(eq)}$
for all time
in the simulation.  This velocity inlet condition amounts to assign
the equilibrium distribution to $f_i$ and it makes  a significant error.
The result is worse than that of pressure inlet condition \cite{bob}.

\subsection{Boundary Conditions for the Modified Incompressible Model d2q9i}

The velocity wall boundary condition and flow boundary conditions for d2q9i
are similar to
that of d2q9. It is from equations $\sum_{i=0}^8 f_i = \rho$ and
$\sum_{i=1}^8 {\bf e}_i f_i = {\bf v}$ and hence some modifications are
needed as follows:
\begin{itemize}
\item In wall boundary condition, Eq.~(\ref{eq:rhod}) is replaced by
\begin{equation}
 \rho = v_y - [f_0+f_1+f_3+2(f_4+f_7+f_8)].
\label{eq:rhodi}
\end{equation}
\par
and in Eq.~(\ref{eq:f256}), $\rho u_x, \rho u_y$ are replaced by
$v_x, v_y$ respectively.
\item In pressure flow boundary condition, Eq.~(\ref{eq:u0d}) is replaced by
\begin{equation}
 v_x =  \rho - [f_0+f_2+f_4+2(f_3+f_6+f_7)] ,
\label{eq:u0di}
\end{equation}
and in Eq.~(\ref{eq:f158}), $\rho_{in} u_x $ is replaced by
$v_x $.
\item Similar replacement in velocity flow boundary condition,
Eq.~(\ref{eq:bc}).

\end{itemize}

\section{Numerical Results and Discussion }

We report and discuss  the numerical results for Poiseuille flow with
pressure (density) or velocity flow boundary condition. The simulation
is performed on both models d2q9 and d2q9i.
The main result in the simulation of d2q9i is the achievement of machine
accuracy.
The main result in the simulation of d2q9 is the achievement of
second-order accuracy of the boundary conditions.
The width of the channel is assumed to be $2 L = 2$.
We use $nx, ny$ lattice nodes on the $x-$ and $y-$directions, thus,
$\delta = 2/(ny-1)$.
The initial condition is to assign $f_i = f_i^{(eq)}$ computed using
a constant density $\rho_0$, and zero velocities.
The steady-state is reached if
 \[ \frac{ \sum_{i} \sum_{j} | u_x(i,j, t+\delta) - u_x(i,j,t)|
 + | u_y(i,j, t+\delta) - u_y(i,j,t)|}
 { \sum_{i} \sum_{j} |  u_x(i,j,t)|
 + |  u_y(i,j,t)|}  \leq  \delta \cdot Tol .  \]
For model d2q9i, $u_x,u_y$ are replaced by $v_x, v_y$. $Tol$ is a tolerance
usually set to $10^{-10}$.
On the wall, boundary condition discussed in section 2.2
is used to make  non-slip boundaries.

We also define a L1 error as:
\begin{equation}
 err_1 \equiv  \frac{ \sum_{i} \sum_{j} | u_x^t(i,j) - u_x(i,j)|
 + | u_y^t(i,j) - u_y(i,j)|}
 { \sum_{i} \sum_{j} |  u_x^t(i,j)|
 + |  u_y^t(i,j)|} ,
\label{eq:err1}
\end{equation}
where $u_x^t, u_y^t$ is the analytical velocity.

\subsection{Results of Model d2q9i}

For model d2q9i, we carried out
simulations with a variety of Re,  $nx, ny, u_0$ ($u_0$ is the peak
velocity in the channel) using the pressure or velocity flow
boundary condition.
The range of Re is from 0.0001 to 30.0;
the range of $\tau $ is from 0.56 to 20.0 and
the range of $u_0 $ is from 0.001 to 0.4;
the largest density difference simulated (not the limit)
is $\rho_{in} = 5.6, \ \rho_{out}=4.4$ with $nx=5, ny=3$ corresponding to
a pressure gradient of $G=0.1$.
The magnitude of average density
$\rho_0$ is irrelevant for the simulation \cite{zoui}.

For all cases where the simulation
is stable, the steady-state  velocity and density show:
\begin{itemize}
\item The velocity field $v_x$ is accurate up to machine accuracy
compared to the analytical solution in Eq.~(\ref{eq:pois}),
$v_y$ is very small with maximum of $|v_y|$ in the whole region being
in the order
of $10^{-12}$. For example,  for
$nx = 5, ny = 3, u_0= 0.1, \tau=0.56, $ Re =10,  the relative L1 error
of ${\bf v}$ in the whole flow region is
$0.485 \cdot 10^{-10}$.
\item The density is uniform in the cross channel direction, and linear in
the flow direction. The density difference $\rho(i+1,j) - \rho(i,j)$
is a constant through the flow region,
its value is equal (up to machine accuracy) to the analytical
value set by the  constant pressure gradient.
\item Velocity $v_x$ is uniform in the $x-$direction, the results are
the same for different $nx$.
\end{itemize}
If the computed velocity were plotted with the analytical velocity,
there would be  no difference to naked-eyes.

It is also noticed that with pressure (density) gradient to drive
Poiseuille flow, the maximum Reynolds
number which makes the simulation stable is far less than that with
external forcing.  For $ny = 5$, the maximum Re is about 30.
Refinement of mesh can increase the maximum Re.

When the pressure flow boundary condition is replaced by the method in
\cite{bob},
machine accuracy can no longer be obtained,
 for a simple case
$nx=17, ny= 9, u_0= 0.03542, \tau=0.67, $ Re =5, with a  moderate
pressure gradient of $G=0.001004$, $\rho_{in} = 5.006, \rho_{out} = 4.994$,
  the relative L1 error of ${\bf v}$ in the whole flow region is
$0.1824 \cdot 10^{-2}$, with  maximum of $|v_y|$  being
$0.1364  \cdot 10^{-3}$, not very small
compared to $u_0$.
The density difference  $(\rho(i+1,j) - \rho(i,j))/\delta$
is no longer a constant, its range is from  -0.002094 to -0.003872
(the analytical value is $-0.003012 = -G /c_s^2= -3G$).
The result indicates that the pressure or velocity flow boundary condition
proposed in this paper is a clear improvement of that in \cite{bob}.

Similar results of machine accuracy
are obtained by specifying the analytical velocity profile
given in Eq.~(\ref{eq:pois})
at inlet and pressure (density) at outlet by using the flow boundary
conditions in this paper.  In the case, there is a uniform
pressure (density) difference in the region. The value of the density
difference
depends on $u_0$ and the outlet density.

\subsection{Results of Model d2q9}

Since the ordinary LBGK model  d2q9 is still widely used
for simulations.  It is worthwhile to do some
simulations with d2q9 with our flow boundary conditions and show that
they give a second-order accuracy.
We use d2q9 to Poiseuille flow with pressure or velocity flow boundary
condition.  Since as $\delta \rightarrow 0$ the computed solution
does not approach the analytical solution, we will use the result
of the finest mesh as the exact solution to compute the error.
Simulations with successively doubled lattice steps  are carried out to
observe the convergence.
The example uses fixed Re = 10, $u_0=0.1, \rho_0 = 5$.
The pressure gradient $G$ in Eq.~(\ref{eq:pois})
and then pressure (density) at inlet/outlet can be
obtained as
$G=0.02$ and $\rho_{in}=5.12, \rho_{out}=4.88$ respectively
to be used in the pressure (density) flow boundary condition.
For the velocity flow boundary condition, the analytical velocities in
Eq.~(\ref{eq:pois}) are used at inlet, and $\rho_0$ is specified at outlet.
Thus, the results of the pressure flow boundary condition and the
velocity flow
boundary condition are similar but not identical.
Define $lx=nx-1, ly=ny-1$ to represent the number of lattice steps in
$x-$ and $y-$directions, we use $lx = 4,8,16,32,64,128,256,
ly = 2,4,8,16,32,64,128$ respectively to do the
simulation. The L1 error is defined in Eq.~(\ref{eq:err1}) with
$u_x^t, u_y^t$ being replaced by the computed velocities of $lx=256, ly=128$.
$\tau$ has to be changed as $lx, ly$ are changed  to keep the same Re, $u_0$
(values of $\tau$ are included in Table~I).
The convergence result is summerized in Table~I.
The case of $lx=4, ly=2$
was not shown in Table I, because the simulation is unstable
for the velocity inlet condition.
The ratio of two consecutive  L1 errors is also shown.
The ratio is approximately equal to 4, indicating a second-order accuracy.
For all runs, it is also observed that
\begin{itemize}
\item Velocity $u_x$ is monotonically increasing in the $x-$direction,
the result is
not sensitive to  the value of  $lx$, we usually take $lx=2 \ ly$.
If the
pressure boundary condition in \cite{bob} is used, $u_x$ on the centerline
of the channel decreases first, then increases, and decreases again
near the outlet. This behavior deviates from the macroscopic continuity
equation of LBGK $\partial_x (\rho u_x) = 0$ in the case, indicating that
errors are introduced by the pressure boundary condition in \cite{bob}.
An example of centerline  velocity $u_x$  as a function of $x$ is presented
in Fig.~4.
\item  $u_y$ is very small compared to $u_0$,
with maximum of $|u_y|$ being approximately
of $10^{-6} \cdot u_0$ in typical  cases.
If the pressure boundary  condition in \cite{bob} is used, maximum value of
$|u_y|$ is like  $10^{-3} \cdot u_0$ typically.
\item The density is uniform in the cross channel direction, and linear in
the flow direction. The density difference $\rho(i+1,j) - \rho(i,j)$
is almost a constant through the flow region, its value is equal to the
analytical
value with a fluctuation   of
less than 0.03~\% in a worst case observed with Re= 5.
If the pressure boundary condition in \cite{bob} is used,
The fluctuation of density difference $\rho(i+1,j) - \rho(i,j)$
from the analytical value reaches  20~\% for the same case mentioned above with
Re = 5.
\end{itemize}

In the case where the density gradient is small, the computed velocity profile
are close to the analytical velocity profile of Poiseuille flow.
We present an example here with
$nx =9, ny = 5$, Re=0.8333, $u_0=0.008333$ , $G = 0.001667$,
$\rho_{in} = 5.01, \rho_{out} = 4.99$ with both pressure (density) flow
boundary
conditions in this paper and in \cite{bob}.
Fig.~2 shows the velocity $u_x$ as a function of $y$ at
$i=5$ (the middle section of the channel),
Fig.~3 shows the centerline density profile along the $x$ direction, and
Fig.~4 shows the centerline $u_x$ along the $x$ direction.
the solid line represents the corresponding analytical
solution and the symbols $\diamond, +$ represent the computed solutions
with the pressure boundary conditions in this paper and in \cite{bob}
respectively.
Both computed velocities $u_x$ at the mid-channel
has no difference to the analytical solution
to naked eyes (the relative error are of order $10^{-3}$ for both
pressure boundary  conditions).
The computed centerline  density with our method looks identical to the
analytical solution
(given as
a linear function crossing  $\rho_{in}$ and $\rho_{out}$  at inlet and outlet
respectively),
while  the centerline density with the method in \cite{bob}
has a discernible difference with the linear function especially near the
inlet and outlet.
The centerline $u_x$ with our method is monotonically increasing,
the behavior is
consistent with the continuity equation of LBGK $\partial_x (\rho u_x) = 0$
in the case,
while the centerline $u_x$ with the method in \cite{bob} has a behavior
inconsistent with the continuity equation near the inlet and outlet.
This again shows an clear
improvement of our method to that in \cite{bob}.

\section{Flow Boundary Conditions and
Preliminary Results  for the 3-D 15-velocity
LBGK Model}

Since 3-D model is needed in practical problems, we give a discussion
of the pressure or velocity flow boundary condition for the 3-D
15-velocity LBGK model (d3q15)
and present a brief statement about its simulation results.
The model is based on the LBGK equation Eq.~(\ref{eq:lbgk})
with $i = 0, 1, \cdots, 14$, where ${\bf e}_i, i = 0, 1, \cdots, 14$ are the
column vectors of the following matrix:
\[ E =  \left[ \begin{array}{rrrrrrrrrrrrrrr}
  0 &1 &-1 &0 &0 &0 &0 &1 &-1 &1 &-1 &1 &-1 &1 &-1       \\
  0 &0 &0 &1 &-1 &0 &0 &1 &-1 &1 &-1 &-1 &1 &-1 &1       \\
  0 &0 &0 &0 & 0 &1 &-1 &1 &-1 &-1 &1 &1 &-1 &-1 &1
         \end{array}    \right]               \]
and ${\bf e}_i, i = 1, \cdots, 6$  are clasified as type I,
${\bf e}_i, i = 7, \cdots, 14$  are clasified as type II.
The density per node, $ \rho $, and the macroscopic flow velocity,
 ${\bf u} = (u_x, u_y, u_z)$,
are defined in terms of the particle distribution function by
\begin{equation}
 \sum_{i=0}^{14} f_i=\rho,  \ \ \ \
 \sum_{i=1}^{14} f_i {\bf e}_i =\rho {\bf u}.
\label{eq:3ddens}
\end{equation}
The equilibrium can be chosen as:
\begin{eqnarray}
f_{0}^{(eq)}&=&\frac{1}{8}\rho- \frac 13 \rho {\bf u}\cdot{\bf u} ,
\mbox{\hspace{1.65 in}}
\nonumber\\
f_{i}^{(eq)}&=&\frac{1}{8}\rho +\frac 13 \rho{\bf e}_{i}\cdot
{\bf u}+ \frac{1}{2}\rho({\bf e}_{i}\cdot {\bf u})^2-\frac{1}{6}\rho
{\bf u}\cdot {\bf u} ,\;\; i =  I \nonumber \\
f_{i}^{(eq)}&=&\frac{1}{64}\rho +\frac{1}{24} \rho{\bf e}_{i}\cdot
{\bf u}+ \frac{1}{16}\rho({\bf e}_{i}\cdot {\bf u})^2-\frac{1}{48}\rho
{\bf u}\cdot {\bf u}  .\;\; i =  II
\label{eq:3dequil}
\end{eqnarray}

Since we will use this model for 2-D simulation in this paper, it is
clear to give a projection of the velocities  in the $xz$ plane
as shown in Fig.~1. The $y-$axis is pointing into the paper, so are
velocity directions 3,7,9,12,14, while the
velocity directions 4,8,10,11,13 are pointing out (velocity
directions  3,4 have
a projection at the center and are not shown in the figure).
The flow direction is still $x$, and the cross channel direction is $z$.
The macroscopic equations of the model is the same as
Eqs.~(\ref{eq:cont},\ref{eq:momen}) with $c_s^2 = 3/8$, and
$\nu = (2 \tau -1)\delta /6$.

The velocity wall
boundary condition proposed in \cite{he2} has the following version
for the model d3q15:
take the case of a bottom node (wall node) as shown in Fig.~1, the wall
is on $xy$ plane.
After streaming, $f_i, (i = 0,1,2,3,4,6,8,9,12,13)$ are known,
Suppose that $u_x, u_y, u_z$ are specified on the wall, we need to
determine $f_i, i = 5,7,10,11,14$
 and $\rho$ from Eqs.~(\ref{eq:3ddens}).
Similar to the derivation in d2q9, $\rho$ is determined by a consistency
condition as:
\begin{equation}
 \rho =  \frac{1}{1-u_z} [f_0+f_1+f_2+f_3+f_4+2(f_6+f_8+f_9+f_{12}+f_{13})].
\label{eq:3drhod}
\end{equation}
The expression of $z-$ momentum gives:
\begin{equation}
 f_5+f_7+f_{10}+f_{11}+f_{14} =
 \rho u_z + (f_6+f_8+f_9+f_{12}+f_{13})  ,
\label{eq:3duz}
\end{equation}
If we use  bounce-back rule for the
non-equilibrium part of the particle distribution
$ f_i, (i=5,7,10,11,14)$ to set
\begin{eqnarray}
f_i = f_{i+1} + (f_i^{(eq)} - f_{i+1}^{(eq)}) , \; i = 5,7,11 \nonumber \\
f_i = f_{i-1} + (f_i^{(eq)} - f_{i-1}^{(eq)}) , \; i = 10,14
\label{eq:bbn}
\end{eqnarray}
then Eq.~(\ref{eq:3duz}) is satisfied, and all $f_i$ are defined.
In order to get the correct
$x-, y-$momenta,  we further fix this $f_5$
(bounce-back of non-equilibrium $f_i$ in the normal direction)
and modify
$f_7, f_{10}, f_{11}, f_{14}$  as in \cite{bob}:
\begin{equation}
  f_i \leftarrow f_i + \frac 14 e_{ix}\delta_x +\frac 14 e_{iy}
\delta_y . \;\; i = 7, 10, 11, 14
\label{eq:bbm}
\end{equation}
This modification  leaves $z-$ momentum unchanged   but adds
$\delta_x, \delta_y$ to the $x-, y-$momenta respectively. A suitable
choice of $\delta_x$ and $\delta_y$ then gives the correct
$x-, y-$momenta. Finally, we find:
\begin{eqnarray}
 f_5&=&f_6 +  \frac 23 \rho u_z ,  \nonumber \\
 f_i&=&f_{j} +  \frac{1}{12} \rho u_z + \frac 14 [e_{ix}(\rho u_x-f_1+f_2) +
 e_{iy}(\rho u_y-f_3+f_4) ] ,
\label{eq:3dbc}
\end{eqnarray}
where $j$ is the index corresponding to
${\bf e}_j = -{\bf e}_i$ (e.g., $j=8$ for $i=7$ and $j=9$ for $i=10$).
In the case of non-slip boundary, this boundary condition is reduced to
that in \cite{bob}.

The derivation of pressure (density) flow boundary condition uses a similar
way as for velocity wall boundary condition.
Suppose the boundary (take the case of inlet in Fig.~1)
 is on $yz$ plane with specified $\rho_{in}$ and
$u_y=u_z =0$.  After streaming,  we need to determine
$u_x$ and $f_i, (i = 1,7,9,11,13)$ from Eq.~(\ref{eq:3ddens}).
The consistency condition gives:
\begin{equation}
 \rho_{in} u_x = \rho_{in} - [f_0+f_3+f_4+f_5+f_6+2(f_2+f_8+f_{10}+f_{12}+
f_{14})  ] ,
\label{eq:3din1}
\end{equation}
which determines  $u_x$ at inlet, using a similar procedure as in deriving
the boundary condition, we find:
\begin{eqnarray}
 f_1&=&f_2 +  \frac 23 \rho_{in} u_x ,  \nonumber \\
 f_i&=&f_j +  \frac{1}{12} \rho_{in} u_x - \frac 14 [e_{iy}(f_3-f_4) +
 e_{iz}(f_5-f_6) ] , \;\; i = 7,9,11,13
\label{eq:3din2}
\end{eqnarray}
where $j$ direction is opposite to $i$ direction.

Same procedure as in d2q9 is used at the inlet bottom node (non-slip) to
derive:
\begin{eqnarray}
 f_1&=&f_2 , \;\; f_5 = f_6 ,  \nonumber \\
 f_7&=&f_8 - \frac 12 (f_3-f_4) ,\;\;
 f_{11} = f_{12} + \frac 12 (f_3-f_4) , \nonumber \\
 f_9&=&f_{10}=f_{13}=f_{14}  \nonumber \\
    &=& \frac 14 [\rho_{in} -
   (f_0+f_1+f_2+f_3+f_4+f_5+f_6+f_7+f_8+f_{11}+f_{12})] ,
\label{eq:3din3}
\end{eqnarray}
and similar results for inlet top and outlet condition.

The pressure (density) boundary condition in \cite{bob} is specified
as post-streaming rule (take the inlet as an example):
\begin{eqnarray}
 f_i({\bf x}, t)= f_i^+({\bf x}, t-\delta) -
 (f_j({\bf x}, t)-f_j^+({\bf x}, t-\delta) ) , \;\; i = 1,7,9,11,13
\label{eq:3din4}
\end{eqnarray}
where $f_i^+({\bf x}, t-\delta)$ is the distribution functions at previous
time step after collision, $f_j$ is along ${\bf e}_j$ with
${\bf e}_j = {\bf e}_i - 2 {\bf e}_n$ (the inner normal
${\bf e}_n = {\bf e}_1$ in the case).
Thus, $j=2,14,12,10,8$ for $i=1,7,9,11,13$ respectively.
Then $\rho$ is set to $\rho_{in}$ and $u_y, u_z$ are set to zero to compute
$f_i^{(eq)}$. Again, the density and $z-$ momentum from $f_i$ may not be
correct, indicating some inconsistency.

The velocity flow boundary  condition, which can be viewed
simply a velocity wall boundary condition, can be derived similarly.
But the
corner node with non-slip condition needs some treatment as in the d2q9 case.
The details are easy to work out and omitted here.

Simple modifications are used to derive wall boundary condition, pressure
(density) or velocity flow boundary condition for the improved incompressible
model d3q15i, which is the counterpart of d2q9i in 3-D case.

Simulations on plane Poiseuille flow are performed on d3q15 and d3q15i
using the pressure or velocity flow boundary condition.
The only difference with
2-D simulations is periodic condition is used on $y-$direction, and
initial condition is uniform in $y$ direction with zero velocity.
It is found that the
result is uniform in the $y$-direction and is independent of the number
of nodes in $y-$direction. $u_y$ is identically zero (in the order of
$10^{-16}$ because of round-off error).
 The results are very similar, although not
identical, to the results of d2q9, d2q9i.
Again, d3q15i with our boundary condition and pressure (density) flow boundary
condition gives results with machine accuracy, showing a clear improvement
over the pressure boundary condition in \cite{bob}.

\section{Acknowledgments}

Discussions with R. Maier, R. Bernard  are appreciated.
Q.Z. would like to thank the Associated Western Universities  Inc.
for providing a fellowship and to thank G. Doolen and S. Chen
for helping to arrange his visit to
 the Los Alamos National Lab.

\vfill\eject


\newpage

 Table I. L1 relative error of velocities as $lx,ly$ are doubled, Re=10,
$u_0=0.1$ with our pressure or velocity flow boundary condition.
Error are from comparison with the velocities of $lx=256, ly=128$
($\tau= 4.34$). The symbol (-2) represents $10^{-2}$.
Ratio of two consecutive L1 errors is also shown.   \\

\begin{tabular}{|c|l|l|l|l|l|l|}  \hline
&lx              &8 &16   &32  &64  &128     \\
inlet &ly              &4 &8 &16   &32  &64       \\
condition &$\tau$     &0.62  &0.74 &0.98 &1.46  &2.42    \\  \hline
pressure &L1 error &0.1049(-2)&0.2522(-3)&0.6135(-4) &0.1458(-4)
&0.2915(-5)    \\
(density)&ratio  &4.159 &4.110 &4.208 &5.003 &  \\
 \hline
velocity &L1 error &0.2301(-3) &0.4882(-4)&0.1167(-4) &0.2774(-5)
         &0.5582(-6)    \\
&ratio   &4.713 &4.183 &4.207 &4.970 &  \\
 \hline
\end{tabular}

\vspace{4mm}

\newpage

\section{Figure Caption}

Fig.~1, Schematic plot of velocity directions of the 2-D (d2q9) model
and projection of 3-D (d3q15) model in a channel.
In the 3-D model, The $y-$axis is pointing into the paper, so are
velocity directions 3,7,9,12,14, while the
velocity directions 4,8,10,11,13 are pointing out (velocity directions
 3,4 have
a projection at the center and are not shown in the figure).

\vspace{1in}

Fig.~2, Velocity $u_x$ as a function of $y$ at
$i=5$ (the middle section of the channel) for the case
$nx =9, ny = 5$, Re=0.8333, $u_0=0.008333$ , $G = 0.001667$,
$\rho_{in} = 5.01, \rho_{out} = 4.99$ with pressure (density) flow boundary
conditions.
The solid line represents the corresponding analytical
solution of Poiseuille flow. The symbols $\diamond, +$ represent the
computed solutions
with the pressure flow boundary conditions in this paper and
in \cite{bob} respectively.

\vspace{1in}

Fig.~3,  Centerline density profile along the $x$ direction
for the case $nx =9, ny = 5$, Re=0.8333, $u_0=0.008333$ , $G = 0.001667$,
$\rho_{in} = 5.01, \rho_{out} = 4.99$ with pressure (density) flow boundary
condition.
The solid line represents the corresponding  analytical
solution
(a linear function crossing  $\rho_{in}$ and $\rho_{out}$  at inlet and outlet
respectively).
The symbols $\diamond, +$ represent the computed solutions
with the pressure flow boundary conditions in this paper and in \cite{bob}
respectively.

\vspace{1in}

Fig.~4,  Centerline $u_x$ along the $x$ direction
for the case $nx =9, ny = 5$, Re=0.8333, $u_0=0.008333$ , $G = 0.001667$,
$\rho_{in} = 5.01, \rho_{out} = 4.99$ with pressure (density) flow boundary
condition.
The solid line represent the analytical solution of Poiseuille flow.
The symbols $\diamond, +$ represent the computed solutions
with the pressure flow boundary conditions in this paper and in \cite{bob}
respectively.

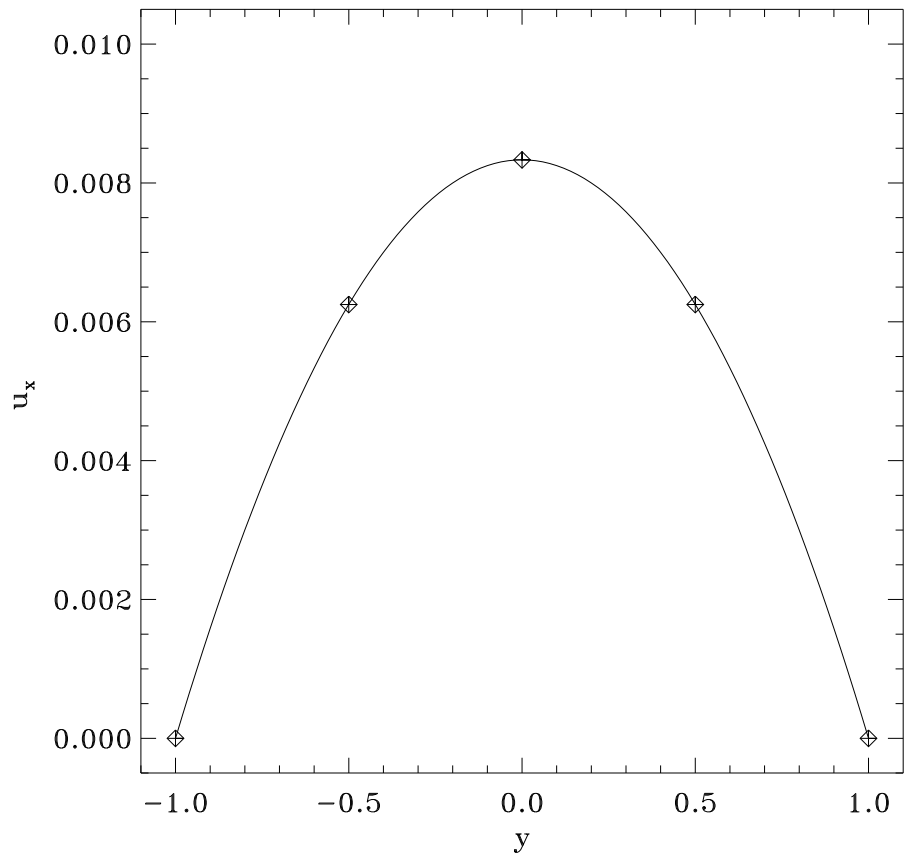
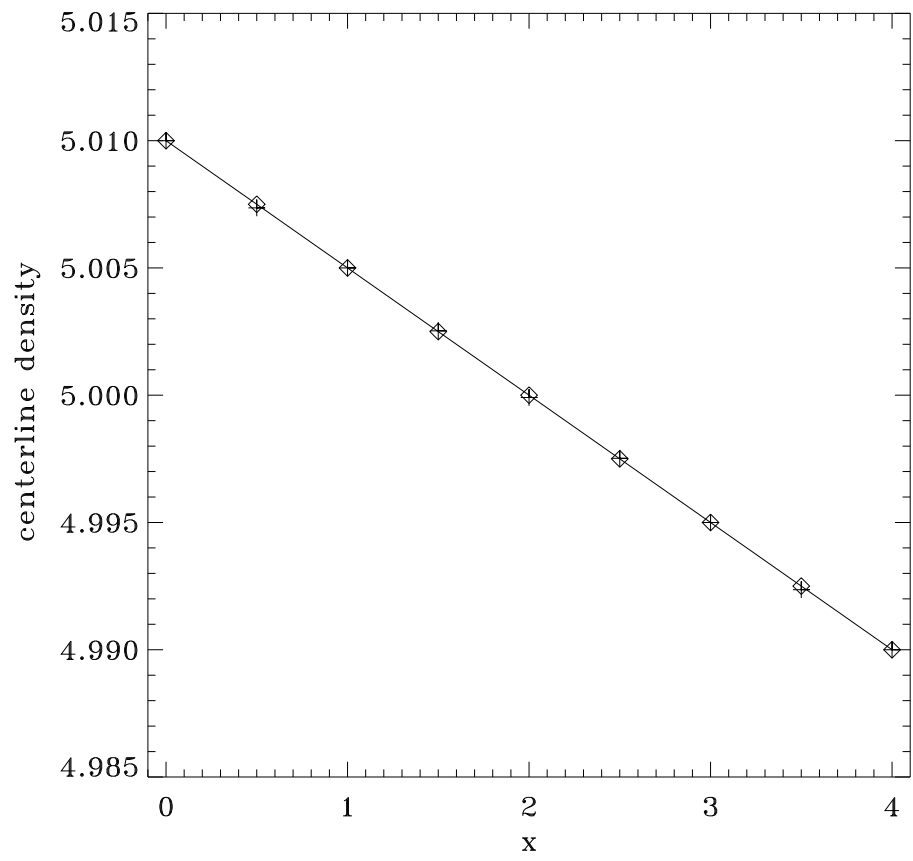
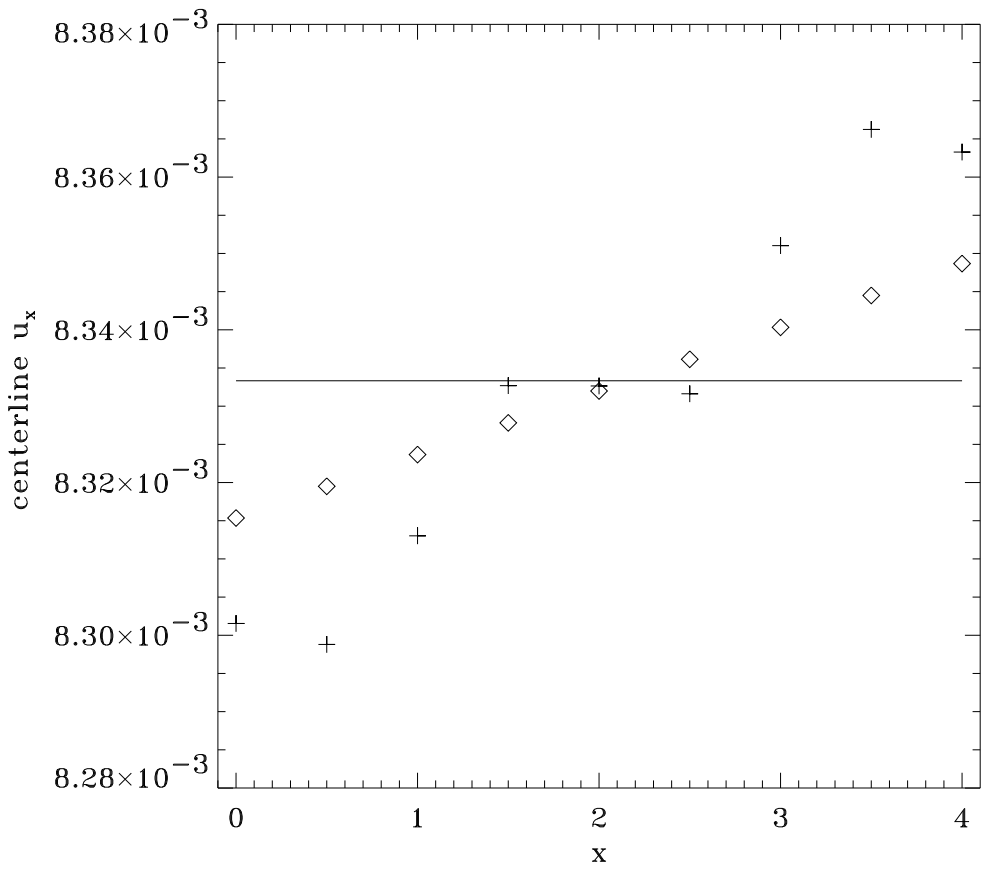
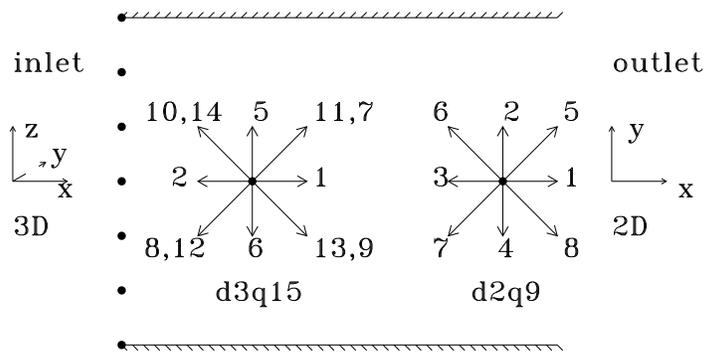
\end{document}